\documentclass{elsarticle}
\usepackage{epsfig}
\usepackage{amsmath}
\usepackage{amssymb}
\usepackage{dcolumn}
\usepackage{graphicx}
\usepackage{dcolumn}
\usepackage{graphicx}
\usepackage{natbib}
\bibpunct{(}{)}{;}{a}{}{,}

\begin{document}
\begin{frontmatter}

\title{Extreme protraction for low grade gliomas: Theoretical proof of concept of a novel therapeutical strategy}

\author{V\'{\i}ctor M. P\'erez-Garc\'{\i}a}


\address{Departamento de Matem\'aticas, E. T. S. I. Industriales and Instituto de Matem\'atica Aplicada a la Ciencia y la Ingenier\'{\i}a, Universidad de Castilla-La Mancha, \\ 13071 Ciudad Real, Spain.}

\date{\today}

\begin{abstract}

Grade II gliomas are slowly growing primary brain tumors that affect mostly young patients and become fatal after a few years. Current clinical handling includes surgery as first line treatment. Cytotoxic therapies (radiotherapy RT or chemotherapy QT) are used initially only for patients having a bad prognosis. Therapies are administered following the 'maximum dose in minimum time' principle, what is the same schedule used for high grade brain tumors. Using mathematical models describing the growth of these tumors in response to radiotherapy, we find that a extreme protraction therapeutical strategy, i.e. enlarging substantially the time interval between  RT fractions, may lead to a better tumor control. 
Explicit formulas are found providing the optimal spacing between doses in a very good agreement with the simulations of the full three-dimensional mathematical model approximating the tumor 
spatio-temporal dynamics.
This idea, although breaking the well-stablished paradigm, has biological meaning since in these slowly growing tumors it may be more favourable to treat the tumor as the different tumor subpopulations move to more sensitive phases of the cell cycle.
\end{abstract}




\end{frontmatter}

\section{Introduction}

Gliomas as a group are the most frequent type of primary brain tumors. With few unfrequent exceptions aside, these tumors remain a challenge for medicine since therapies cannot eradicate them due to their infiltrative nature. Patients diagnosed with gliomas typically die because of the complications related to the tumor evolution.

Low grade glioma (LGG) is a term used to describe WHO grade II primary brain tumors of astrocytic and/or oligodendroglial origin \citep{WHO}. They represent a subgroup of gliomas with moderate incidence that are diagnosed mostly in young adults.

These tumors have a slow growth but are highly infiltrative and generally incurable, the median survival time being of about $5$ years \citep{Pignatti2002,Pouratian2010}.
While some patients present with easy to control seizures and remain stable for many years, others undergo soon the so called malignant transformation and progress rapidly, with increasing neurological symptoms, to a higher-grade tumor.

Management of LGG is controversial because many of these patients present few, if any, neurological symptoms. Recent evidences support that the early use of surgery  results in a better outcome than the historically used watch and wait approach \citep{Grier2006,Smith2008,Jakola2012}.

The decision as to whether a patient with LGG should receive resection, radiation therapy (RT), or chemotherapy is based on a number of factors including age, performance status, location of tumor, and patient preference \citep{Ruiz2009,Pouratian2010}. Since LGGs are such a heterogeneous group of tumors with variable natural histories, the risks and benefits of each therapy must be carefully balanced with the data available.

As to RT, the clinical trial by \citet{trial6} showed the advantage of using radiotherapy in addition to surgery. However, the timing of radiotherapy after biopsy or debulking is debated \citep{Chao2006}. It is now well known that immediate radiotherapy after surgery increases the time of response (progression-free survival), but does not seem to improve overall survival \citep{VandenBent2005}.  At the same time the therapy induces serious neurological deficits as a result to normal brain damage.
 Thus, radiotherapy is usually offered to patients with a combination of poor risk factors such as age, sub-total resection, or diffuse astrocytoma pathology \citep{trial5} or those suspicious of having undergone the malignant transformation.

Mathematical modeling has the potential to help in selecting LGGs patients that may benefit from radiotherapy and in developing specific optimal fractionation schemes for selected patient subgroups. The increasing availability of systematic and quantitative measurements of tumor growth rates provides bench examples and typical features of the dynamics providing key information for the development and validation of such models (see e.g. \citet{Pallud2012a,Pallud2012} for LGGs).

Most of the mathematical research on gliomas has been focused on the study of high-grade gliomas \citep{Swanson2003,Clatz2005,Benalli2005,Stamatakos2006,Stamatakos2006b,Fedotov2007,Frieboes2007,Bondiau2008,Swanson2008,Wang2009,Tanaka2009,Eikenberry2009,Roniotis2010,Lowengrub2010,Gu2012,Swanson2011,Alicia2012,Alicia2014,PerezGarcia2014}. Most of these models add different layers of complexity on top of the basic Fisher-Kolmogorov reaction diffusion equation \citep{Murray2007}
to be described in detail later. The role of radiotherapy has also been studied mostly o high-grade tumors \citep{Rockne2010,BondiauRT,Konokoglu2010,Kirkby2010,Michor2014}. As to LGGs there have been very few relevant works using mathematical models to describe the response to RT. \citet{Ribba} developed a model based on ordinary differential equations describing the response of low-grade glioma to different therapies with a number of undetermined parameters that can be fit to describe the individual patient's response with a good qualitative agreement. More recently \citet{PerezGarcia2014} have developed a simple spatial model able to describe the known phenomenology of the response of LGGs to RT including the observations from \citet{Pallud2012}. Also \citet{Bratus2014} have found that small variations of the standard dose distributions and/or changes in the fractionation lead only to minor improvements at the best, in agreement with clinical experiences.

However it is not clear from the biological point of view that current scheduling of radiotherapy is optimal considered from a more global point of view. The fact that radiation doses are given in a very short period of time (i.e. one dose per day during five weeks with breaks in the weekends) comes from the idea of reducing the tumor load as much as possible in order to kill every clonogenic cell without allowing the tumor to regrow between doses. While this is a reasonable practice when radiation therapy is used with curative intentions, it is not obvious that an incurable tumor such as a LGG should receive doses in the same way. Indeed LGG grow very slowly, with a very low number of mitoses seen per field what means that only a small fraction of tumor cells is proliferating at a given time and is sensitive to radiation therapy. Thus, it may be  reasonable to enlarge the distance between fractions, i.e. to restort to a protracted therapeutical scheme, to allow for more tumor cells to enter the cell cycle rendering radiation fractions more effective. This idea opens the door to substantially modified schedules that will be studied using mathematical models in this paper.

Our plan in this paper is as follows. First in Sec. \ref{model} we present the mathematical models to be used through the paper. Next in Sec. \ref{results} we  find the optimal therapeutical  protocols and discuss the expected gain as a function of the parameters.  Explicit formulas are found providing the spacing between doses as a function of the biological parameters of the tumor in very good agreement with the results of full three-dimensional simulations. Finally in Sec. \ref{conclusions} we discuss the implications of our results and summarize our conclusions.

\section{Mathematical model}
\label{model}

To fix the mathematical model we have to describe what is: (i) the dynamics of the tumor, (ii) the response of the tumor to radiation, and (iii) the optimization criterion to be used in designing the dose scheduling.

\subsection{Tumor cell dynamics}
\label{dynamics}

For the tumor cell dynamics we will use the simplest model accounting for the growth of a spatio-temporal density $u(x,t)$ of tumor cells in units of a maximal cell number that proliferate with a typical time $1/\rho$ and have a characteristic mobility (diffusion) coefficient $D$, namely the Fisher-Kolmogorov equation \citep{Murray2007}

\begin{subequations}
\label{conc}
\begin{equation}
\frac{\partial u}{\partial t} =  D\Delta u+ \rho (1-u)u, \\
\end{equation}
on a domain $\Omega$ of the brain supplemented with initial data
\begin{equation}
u(x,t_0)  =  u_0(x), \ \ u_0(x) \in C^2(\bar{\Omega}), \label{iu}
\end{equation}
and no-flux boundary conditions
\begin{equation}\label{BC}
\dfrac{\partial u}{\partial \bar{n}}\biggl|_{\partial \Omega} = 0, \end{equation}
\end{subequations}

This model has been extensively used to describe the dynamics of low grade gliomas \citep{Badoual2012,Swanson2007,Swanson2007b,Roniotis2010}. Let us note (see e.g. \citet{ghost}) that
$0 \leq u(x,t) \leq U(t)$ where $U(t)$ is an upper bound for the tumor amplitude solving 
\begin{equation}\label{sim}
\frac{dU}{dt} = \rho (1-U)U,
\end{equation}
with $U(0) = \max_{x_\in \Omega} u(x,0)$. In addition to providing an upper bound for the tumor cell density, Eq. (\ref{sim}) has also been proven to provide the evolution of the tumor amplitude $A(t) = \max_{x \in \Omega} u(x,t)$ with good accuracy in one-dimensional scenarios \citep{CC2012}. Since we will be dealing with low grade gliomas the focus of the therapy will be on limiting disease symptoms and 
delaying the most its transformation into a higher-grade tumor. It has been discussed that high density cell foci may lead to tumor transformation \citep{Swanson2011,PerezGarcia2014}, thus, in this paper we will first focus on understanding and controlling the dynamics of Eq. \eqref{sim} in Secs. \ref{results:amplitude} to \ref{results:examples}. Next, in Sec. \ref{results:3D} we will complete our study with a comparison of our findings with the results for the full dynamics of Eq. (\ref{conc}) in three-dimensional scenarios.

\subsection{Response to RT}
\label{response}

Radiotherapy is given in a short time (typically about 10 minutes) in comparison with typical cellular proliferation times in low-grade gliomas, what allows us to assume the effect of RT to be instantaneous. 

Following the standard practice in radiotherapy
we will assume the damaged fraction of tumor cells  to be given by the classical linear-quadratic (LQ) model \citep{Joiner2009}. Thus for a radiation dose $d_j$ 
given at a time $t_j$, we will take the survival fraction $S_j$, i.e. the fraction of cells that are not lethally damaged to be given by
\begin{equation}\label{LQ}
S_{j}=e^{\displaystyle{-\alpha_t d_j -\beta_t d_j^{2}}}.
\end{equation}
The parameters $\alpha_t$ 
 and $\beta_t$  
 are respectively the linear and quadratic coefficients for \emph{tumor} cell damage of the LQ model.

 The full treatment consists of a total dose $D$ split in a series of $N$ radiation fractions with doses per fraction $\{d_j \}_{j=1,...,N}$
 given at irradiation times $\{ t_j \}_{j=1,...,n}$. The tumor cell density of functionally alive tumor cells $u(x,t)$ at the irradiation times will then satisfy
\begin{eqnarray}\label{discrC}
u(x,t^+_j) & = &  S_{j} u(x,t_j^-),
\end{eqnarray}
what reflects directly on a similar equation for the tumor amplitude $U(t)$: 
\begin{equation}\label{RTcont}
U(t_j^+) = S_j U(t_j^-).
\end{equation}

\subsection{Optimization problem}
\label{Opti}

 Damage to normal tissues caused by radiotherapy can be estimated using  Eq. \eqref{LQ} but taking instead the parameters of the normal tissue $\alpha_h, \beta_h$ with $\alpha_h/\beta_h \simeq  2$ \citep{Wigg}. It is reasonable that any radiotherapy fractionation should keep the damage to the normal tissue equal or below that of the standard fractionation, what can be quantified, given a certain set of doses $(t_j,d_j)$ as (see e.g. \citet{Joiner2009} pp. 114).

\begin{equation}
\label{isoeffect}
E_h  = - \log \left[ \prod_{j=1}^n S_{j} \right] = \alpha_h \left(D + \frac{1}{\alpha_h/\beta_h} \sum_{j=1}^n d^2_j\right).
\end{equation}

In addition, acute tissue reactions and other secondary effects depend: (i) on the total volume irradiated (the so-called volume effect) and (ii) on the maximal dose per fraction $d_*$ used. We will not be interested in this paper on spatial aspects of radiation therapy or other complications \citep{Gong2013} and thus will assume that all tumor cells within the tumor receive the same ammount of radiation.

Mathematical optimization has the potential to help in finding optimal therapeutic schedules, spatial dose distributions, etc. To do so
once a particular tumor evolution model has been chosen and paired with a model for the response to radiation we must choose a cost functional related to the optimization objective.

In the case of high grade brain tumors, where there is no metastatic spreading it has been hypothesized that death occurs in high grade tumors after the tumor reaches a critical diameter of about 6-7 cm \citep{Swanson2008, Wang2009}. However, high grade gliomas are very aggressive tumors that due to the formation of necrotic areas result is a complete loss of functionality in areas affected by macroscopic tumor \citep{PG2011}. 

In LGGs the situation is different. Because of the low density of the tumor the tissue affected is still partially functional and the slow tumor growth allows the brain  to remap continuously the affected functionality into healthy brain areas.
The most harmful effect of the LGG is its potential to transform into a high-grade tumor. So the optimal therapy would be the one delaying the most the malignant transformation of the tumor while at the same time helping to keep the disease symptoms under control. 

One of the driving forces of the malignant transformation is an increase of mutation rates originated by the changes in tumor microenvironment 
 due to the continuous density increase. These include vessel damage, generation of hypoxic foci, the stabilization of hypoxia dependent signaling molecules such as HIF-1$\alpha$ and the increase of genomic instability \citep{Alicia2012,Semenza,Poon,Alicia2014}.

Thus, we will design the therapy to maintain the tumor density below a critical level $U_*$ for the longest time possible, i.e.,  to make the time $T$ such that
\begin{equation}
\label{T2}
 T : \max_{\Omega} u(t, x) \le u_* \equiv U(t) \le U_*
 \end{equation}
 as large as possible. Maximizing $T$ given by Eq. (\ref{T2})  
under the constraints of maximal dose allowed $(d_j \le d_*, \; k = 1,\dots,N)$ and iso-damage to the normal tissue
given by Eq. (\ref{isoeffect}) is a mathematical optimization problem that can be solved for the number of doses $N$, irradiation times $\{t_j\}_{j=1}^N$ and doses $\{d_j\}_{j=1}^N$. In what follows we will refer to the time $T$ given by Eq. (\ref{T2}) as the time to malignant transformation (TMT).

In this paper we will not study the full optimization problem and our focus will be on demonstrating a theoretical proof of concept of the extreme protaction strategy for LGGs, thus grounding mathematically a radically different approach to the planning of radiotherapy for LGGs. 
As such we will fix the number of radiation doses and doses per fraction to be exactly those of the most extended  radiotherapy protocols for these tumors, i.e. take the number of fractions $N=30$ and doses per fraction $d_j = 1.8$ Gy, what keeps the value of the functional $E_h$ in Eq. (\ref{isoeffect}) and satisfies the restriction that $d_j \leq 3.2$ Gy. Our only optimization parameter will be then the time spacing between doses. 

\subsection{Parameter estimation}
\label{sec-parameters}

In the problem to be studied in this paper there remain few biological parameters describing at a macroscopic level the essentials of the LGG dynamics and response to therapy. First, proliferation rates for LGGs have been estimated to be around 0.003 day$^{-1}$ in \citet{Badoual2012} what gives doubling times of the order of one year. From the most indolent tumors to those more aggressive within the class of LGGs the full range may comprise an order of magnitude.  The surviving fraction $S_f$ for doses of 1.8 Gy is harder to estimate. It is known that gliomas are very radioresistant tumors with high surviving fractions in-vitro and mixed response to radiation in-vivo. We will follow the approach of \citet{PerezGarcia2014} that estimate this value from the radiological response fitting the typically observed dynamics of the mean tumor diameter \citep{Pallud2012a} what gives values in the range $S_f \sim 0.8-0.9$. 

Finally, the tumor cell densities leading to relevant symptoms and disease detection ($U_0$) and the threshold density beyond which harmful symptoms appear and/or promoting the malignant transformation ($U_*$), are difficult to estimate. Normal brain tissue has low cellularity and the cell density leading to symptoms is probably dependent on the location of origin of the tumor. However, most LGGs are supratentorial and appear in the white matter. We will take the initial density to be around $U(0) = 0.3$ what means that symptoms arise when 30\% of the space is occupied by tumor cells, a number well beyond the normal physiological value that may be around 10-15\%. As to the maximal tissue density leading to irreversible damage, we will take it to be around $U_* = 0.5-0.6$.

The biological and clinical parameters used through this paper are summarized in Table \ref{parameters}.

\begin{table}[h]
\begin{center}
\label{parameters}
\begin{tabular}{llll}
\hline
Variable  & Description & Value (Units) & References \\ \hline
$\rho$ & Proliferation  & $\sim$ 0.003 day$^{-1}$ &  \citet{Badoual2012} \\ 
          &    rates              &                                                 &                                       \\ \hline
          $d $ & Dose per fraction  & 1.8 Gy &  \citet{trial1} \\ \hline
          $N$ & Number of doses  & 25-30  &    \citet{Soffieti2010} \\ \hline
          SF$_{1.8}$ & Survival fraction   &  $\sim$ 0.8-0.9  &  \citet{PerezGarcia2014}  \\ \hline
      $U(0)$              &     Initial cell density        &   0.3  & Estimated                               \\ \hline
    $U_*$                   & Critical cell density &   0.5-0.6   &  Estimated  \\ \hline
\end{tabular}
\caption{Values of the biological and clinical parameters used in the mathematical model of LGG evolution.}
\end{center}
\end{table}

\section{Results}
\label{results}

\subsection{Equations for the amplitude evolution}
\label{results:amplitude}

Solving explicitly Eq. (\ref{sim}) and using the recursion formula (\ref{RTcont}) we get a recursive equation providing the tumor amplitude after every dose $t_j, d_j$, $U_{k+1} = U(t_{k+1}^+)$
\begin{equation}
\label{recursive}
U_{k+1} = \frac{S_k U_k e^{\rho \left(t_{k+1}-t_k\right)}}{1+ U_k \left[ e^{\rho \left(t_{k+1}-t_k\right)} - 1 \right]}.
\end{equation}
In this paper we will discuss only the case when doses are equispaced
\begin{equation}
 \Delta \equiv t_{j+1}- t_j,
 \end{equation}
 and equal, thus $S_k \equiv S_f$. 
 It is posible to proceed recursively from Eq. (\ref{recursive}) to get 
\begin{equation}\label{Uk}
U_k = \frac{U_0 \left(\alpha S_f\right)^k}{1+ U_0 \left(\alpha-1\right) \sum_{j=0}^{k-1} \alpha^k} = 
\frac{U_0 \left(\alpha S_f\right)^k}{1+ U_0 \left(\alpha-1\right) \left[ \frac{(\alpha S_f)^k-1}{\alpha S_f -1}\right]},
\end{equation}
were $\alpha = \exp \left(\rho \Delta\right)$.

\begin{figure}
\centerline{\epsfig{file=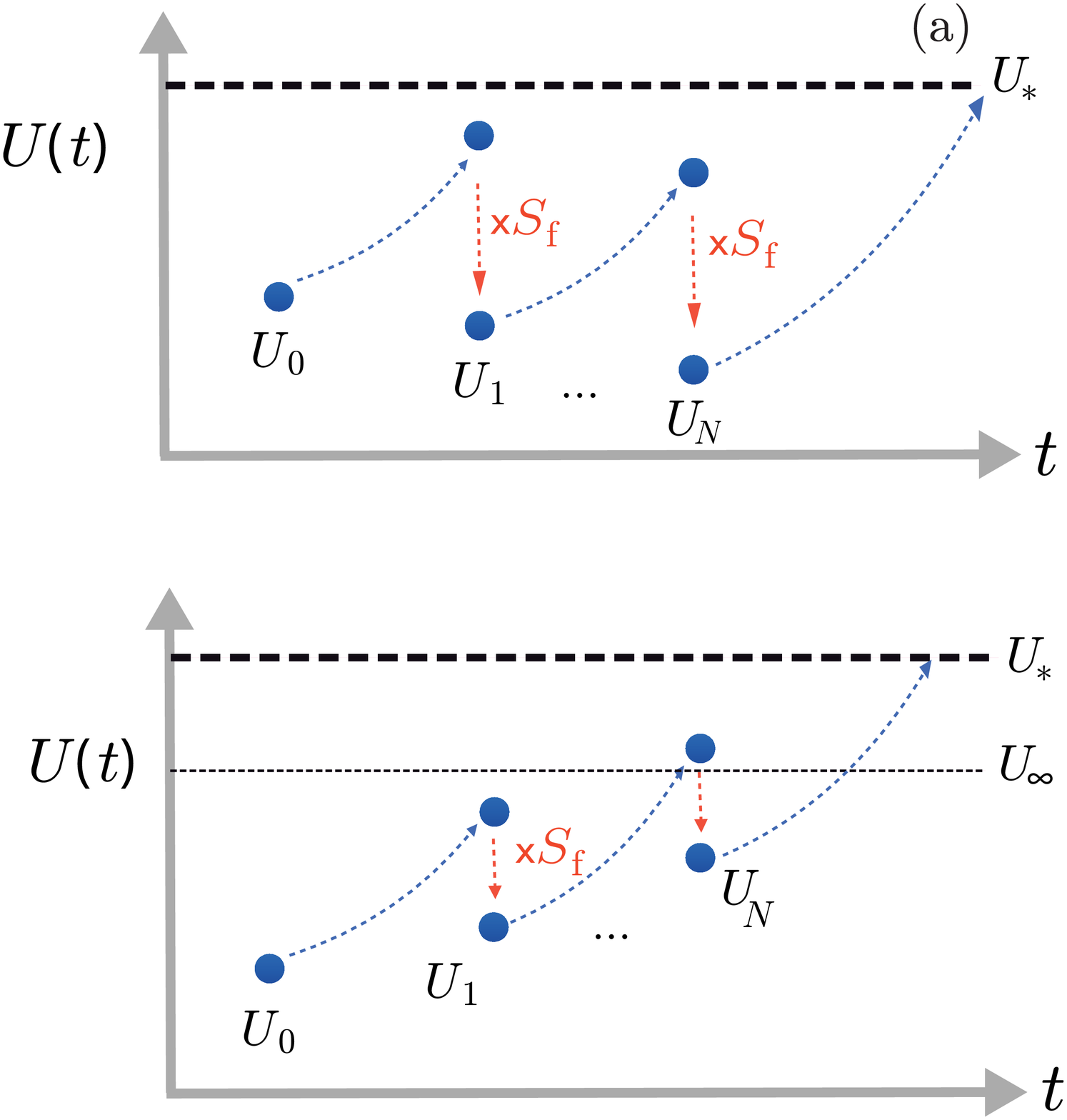,width=8cm}}
\caption{Scenarios for the response of the tumor to radiotherapy depending on the values of $\alpha S_f$. (a) For 
$\alpha S_f < 1$ the therapy provides a net decrease of the tumor amplitude between doses and the maximum is reached after the treatment is completed. (b) For $\alpha S_f > 1$ the tumor continues growing during the therapy. \label{prima}}
\end{figure}

\subsection{Fixed points and types of dynamics}
\label{results:fixedpoints}

Eq. (\ref{recursive}) defines a discrete map whose fixed points are given by the solutions $U_{\infty}$ of the equation
\begin{equation} \label{fp}
U = \frac{S_f U \alpha}{1+(\alpha-1) U}.
\end{equation}
In the biologically relevant regime $0 \leq U \leq 1$, 
Eq. (\ref{fp}) has two solutions: $U=0$ that is globally stable whenever $S_f\alpha < 1$ and $U_{\infty} = (S_f \alpha -1)/(\alpha -1)$ 
that has biological meaning and is stable whenever $S_f \alpha >1$ and $U_{\infty}<1$.  The condition $S_f \alpha = 1$, separates two different types of dynamics: those in which the tumor amplitude decreases through the therapy and then regrows towards $U_*$ ($S_f \alpha >1$  see Fig. \ref{prima}(a)), and those situations 
in which the tumor amplitudes after irradiation continue growing towards $U_{\infty}$ ($S_f \alpha >1$ see Fig. \ref{prima}(b)).
 In the later case, if $S_f U_{\infty}$ (see Fig. \ref{prima}(b)) is larger than the critical amplitude $U_*$ the  criterion is fullfilled before the end of the therapy what is clearly a sub-optimal situation. It is important to note that because of the radioresistance of these tumors $S_f > U_*$. 

\subsection{Optimization problem and its solution}
\label{results:optimization}

Thus, we wish to find, given $\rho, S_f, U_*$, what is the choice of $\Delta$ providing the maximum $T$ given by Eq. (\ref{T2}) under the condition $S_f > U_*$. 
\begin{figure}
\begin{center}
\epsfig{file=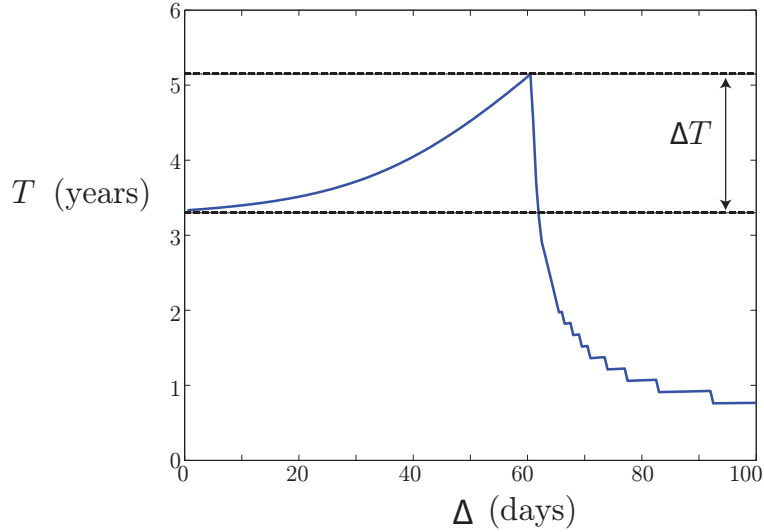,width=10cm}\end{center}
\caption{Time to malignant transformation given by Eq. (\ref{T2}) for different interdose spacings $\Delta$ computed simulating Eq. (\ref{sim}) with doses given at equispaced times according to Eq. (\ref{RTcont}). Parameter values are 
$N=30, S_f = 0.85, U_0 = 0.3, U_* = 0.5$ and $\rho = 0.005$. A maximum value of the time $T$ is obtained for a value of $\Delta$ given by Eq. (\ref{optfin}). Below that value the function $T(\Delta)$ shows a monotonous growth and beyond that the critical $U_*$ is reached before the full treatment is finished, the effective number of doses being smaller than $N=30$. In this case the improvement in time to transformation by choosing the optimal interdose spacing is about $\Delta T = T(\Delta_{\text{opt}}) - T(1) \simeq $ 660 days.
\label{RTeffect}}
\end{figure}
We are interested on the range of values of $\Delta$ for which the 
therapy can be completed before the tumor amplitude reaches $U_*$. Choosing $\Delta$ above that range leads to a suboptimal use of the therapy and to smaller values of the objective function. 
 In the range of interest we can compute explicitly the time $T$ in Eq. (\ref{T2}), that is given by the total time of treatment plus the regrowth time, i.e.
 \begin{equation}\label{TD}
 T(\Delta)  =  N \Delta + \frac{1}{\rho} \log \left[ \frac{U_*(1-U_N)}{U_N(1-U_*)}\right]
 \end{equation}
with $U_N$ given by Eq. (\ref{Uk}) with $k=N$, i.e.
\begin{equation}\label{UN}
U_N = \frac{U_0 \left(\alpha S_f\right)^N}{1+ U_0 \left(\alpha-1\right) \left[ \frac{(\alpha S_f)^N-1}{\alpha S_f -1}\right]}.
\end{equation}

\begin{figure}
\centerline{\epsfig{file=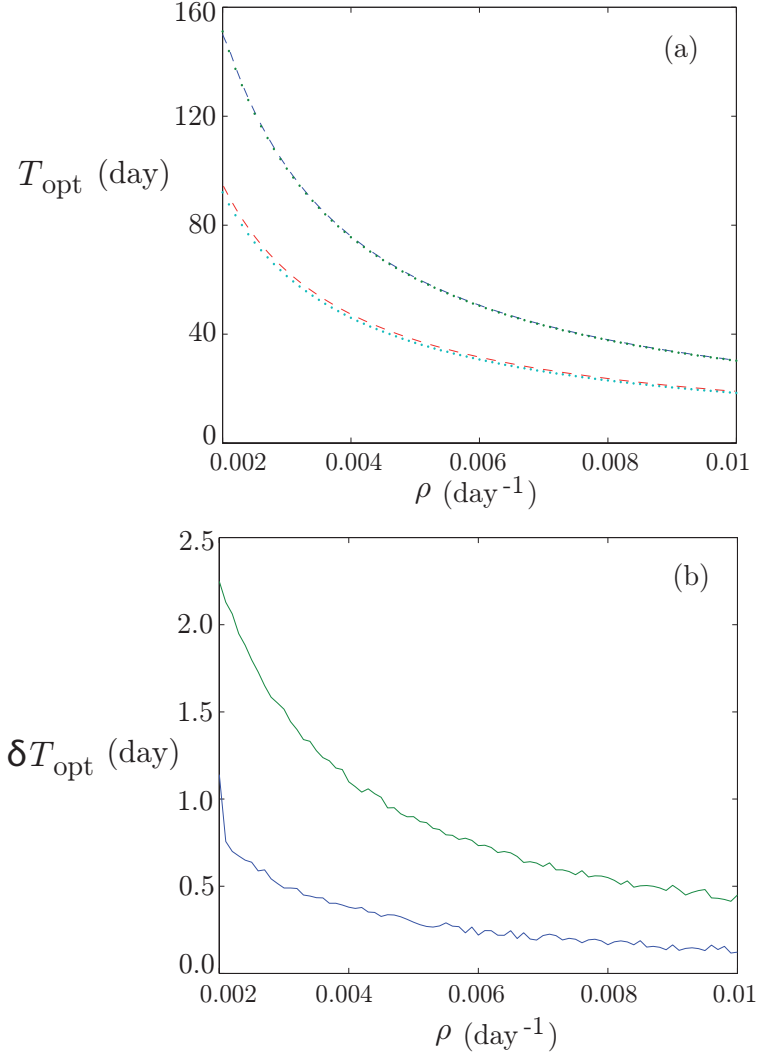,width=10cm}}
\caption{Comparison of the approximations for $T_{\text{opt}}$ obtained from the approximate equation (\ref{optfin}) (dotted lines) and from the exact root of the polinomial (\ref{poli2}) (dashed lines). Parameter values are $U_0 = 0.3, S_f = 0.85, U_* = 0.5$ (upper lines) and $U_0 = 0.3, S_f = 0.9, U_* = 0.45$ (lower lines). Shownare: (a) the values for $T$ as a function of the proliferation $\rho$ and (b)  the differences $\delta T$ between the exact values and the predictions of Eq. (\ref{optfin}) that are shown to be valid to a few percent units, and thus negligible for practical purpouses. \label{segunda}}
\end{figure}

It is a very long but straightforward calculation to prove that 
\begin{equation}
\frac{dT(\Delta)}{d\Delta} > 0,
\end{equation} 
what means that the optimum value for $\Delta$ within this interval is reached exactly on its border. Keep in mind that Eq. (\ref{TD}) is valid as far as $U(t) \leq U_*$. Since $U_k$ gives the amplitude \emph{after} the radiation doses, this means that the condition for this scenario to hold is (cf. Fig. \ref{prima}(b))
\begin{equation}\label{condi}
U_N/S_f \leq U_*
\end{equation}
And the equality is fullfilled for the larger value of $\Delta$ for which the therapy is completed before the tumor reaches the critical value. Fig. \ref{RTeffect} shows an example computed directly from Eqs. (\ref{sim}) with doses given at equispaced times according to Eq. (\ref{RTcont}) showing the existence of a single global maximum in the time to transformation given by Eq. (\ref{T2}). 
We can compute quantitatively the optimum interdose spacing by inserting \eqref{Uk} in Eq. (\ref{condi}) and solving for $\alpha$, i.e. solving the algebraic equation
\begin{equation}\label{poli1}
\frac{U_0}{U_*} \alpha^N S_f^{N-1} = 1 + U_0 (\alpha-1) \frac{(\alpha S_f)^N - 1}{\alpha S_f - 1} 
\end{equation}
Let us define $x = \alpha S_f$, then, we can transform Eq. (\ref{poli1}) into
\begin{multline}\label{poli2}
P(x)  \equiv \left[\frac{U_0}{S_f} \left(\frac{1}{U_*}-1\right)\right] x^{N+1} + \left[U_0\left(1-\frac{1}{U_*S_f}\right)\right] x^N + \\ \left(\frac{U_0}{S_f}-1\right) x + 
\left(1-U_0\right) = 0.
\end{multline}
In the range of parameters of interest ($0 < U_0 < U_* < S_f < 1$) it is simple to prove that Eq. (\ref{poli2}) has at least a root with $x>1$. Moreover we can get an estimate for the value of that root assuming $N \gg 1$ and then we may retain only the leading terms in the polinomial to get
\begin{equation}
\left[\frac{U_0}{S_f} \left(\frac{1}{U_*}-1\right)\right] x^{N+1} + \left[U_0\left(1-\frac{1}{U_*S_f}\right)\right] x^N \approx 0,
\end{equation}
and from here
\begin{equation}
x \approx \frac{1-U_*S_f}{1-U_*},
\end{equation}
what leads to the result
\begin{equation}\label{optfin}
\Delta_{\text{opt}} \approx \frac{1}{\rho}  \log \left(\frac{1/S_f-U_*}{1-U_*}\right),
\end{equation}
that is notably independent of $U_0$ as it should be since after a large ($N$=30) number of doses the dynamics is already close to the fixed point  of the map given by Eq. \eqref{fp} independently of the initial value.

Eq. (\ref{optfin}) provides a simple solution to our optimization problem that is very accurate in the range of parameters of interest as shown in Fig. \ref{segunda}.

\begin{figure}
\begin{center}
\epsfig{file=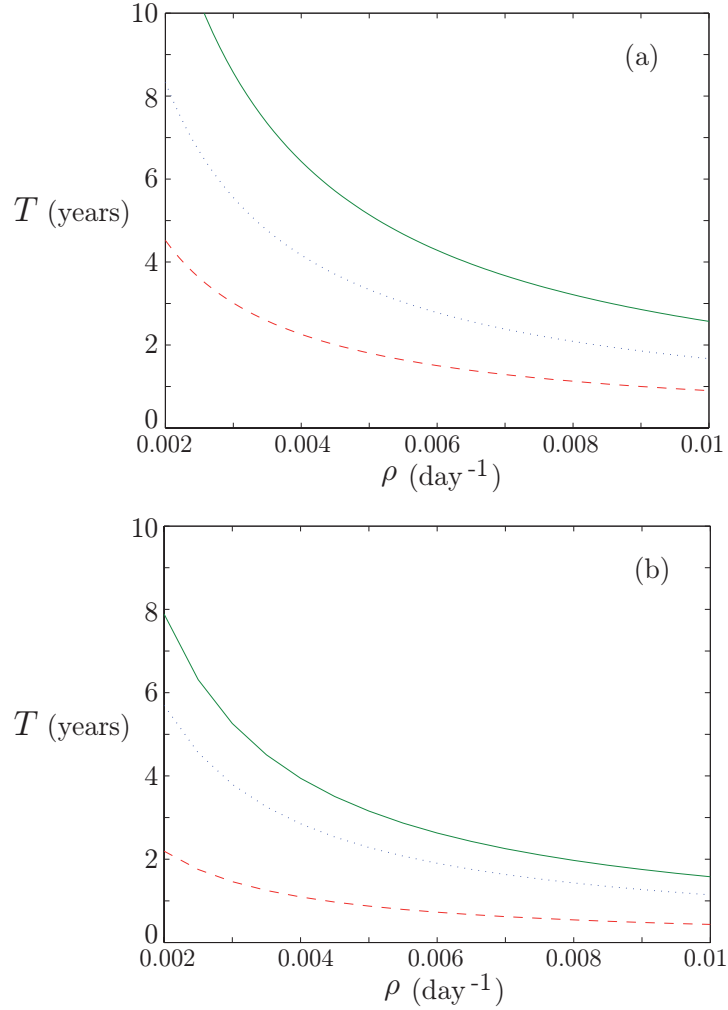,width=9.5cm}
\caption{Comparison of the time to transformation given by Eq. (\ref{TD})  taking the optimal interdose time spacing  $\Delta$ (solid lines) versus the standard choice for $\Delta = 1$ day (dotted lines). The difference between both curves is the estimated improvement because of the choice of the optimal therapy (dashed lines). Shownare the results for $N=30$, and two different sets of parameter values. (a) $S_f=0.85, U_0 = 0.3, U_* = 0.5$, (b) $S_f = 0.9, U_0 = 0.3, U_* = 0.45$. In both cases $T$ is a function of $\rho$ in the range $\rho \in [0.002,0.01].$ \label{comparisonT}}
\end{center}
\end{figure}

\subsection{Examples}
\label{results:examples}

While it is clear that within the framework of our model the choice for the spacing between doses $\Delta_{\text{opt}}$ will lead to a longer time to the malignant transformation than the customary choice $\Delta =1$ day, it is interesting to quantify this improvement in lifetime quantitatively. To do so we have explored the allowed range of parameters extensively and report here a few representative examples. In all the examples to be studied hereafter we have taken the number of fractions to be $N=30$ in agreement with usual clinical practice. 

Let us first take typical values for all tumor parameters. First for the detection density $U_0 = 0.3$ that may be quite typical, next for the transformation density $U_* = 0.5$ and finally an average radiosensitivity of 
$S_f = 0.85$. The corresponding times to transformation following the optimal strategy versus the customary choice of $\Delta =1$ according to Eq. (\ref{TD}) are shown in Fig. \ref{comparisonT}(a). This choice of parameters for average growing tumors ($\rho \approx 0.004$ day$^{-1}$) gives typical times to transformation of about 5 years. In that case the improvement in time to transformation, and correspondingly in survival using the optimal interdose spacing, is about 2.5 years. This result is obtained with $\Delta _{\text{opt}} = 0.30/\rho$ (cf. Eq. (\ref{optfin})), what gives $\Delta _{\text{opt}} \approx 76$ days for $\rho = 0.004$ day$^{-1}$.

Taking a more extreme set of values with a more radioresistant tumor $S_f = 0.9$, and a stricter criterion for transformation $U_* = 0.45$ we get from Eq. (\ref{optfin}) $\Delta _{\text{opt}} = 0.184/\rho$ what leads to 
$\Delta _{\text{opt}} \approx 46$ days for moderately growing tumors ($\rho = 0.004$ day$^{-1}$).  The improvement in transformation time for this set of parameters is shown in Fig. \ref{comparisonT}(b). 

From both examples it is clear that the faster does the tumor grow, the smaller is the gain provided by choosing the optimal interdose spacing. However, for those tumors for which the life expectancy is very bad, even small therapeutical improvements in the range of several months are considered to be significant.

\subsection{Validation on three-dimensional spatiotemporal scenarios}
\label{results:3D}

Through this paper we have taken the tumor compartment dynamics between radiation doses as given by Eq. (\ref{sim}), instead of the more complete model (\ref{conc}). There are many evidences that such approach may be valid ranging from the results of the collective coordinate methods \citep{CC2012} or the fact that it provides a bound for the full dynamics \citep{ghost}. However, Eq. (\ref{sim}) might be an oversimplification of the phenomenon. 

To rule out this possibility in the range of parameters of interest we have compared the results of the predictions of Sec. \ref{results:optimization}
with simulations of the full three-dimensional problem given by Eq. (\ref{conc}) on a cube (i.e. with $\Omega = [-L,L]\times  [-L,L] \times  [-L,L]$). In our simulations we have used second-order algorithms both for space and time. The results have been cross-checked using different temporal and spatial steps and computational domain sizes to avoid spurious discretization effects. 

\begin{figure}
\begin{center}
\epsfig{file=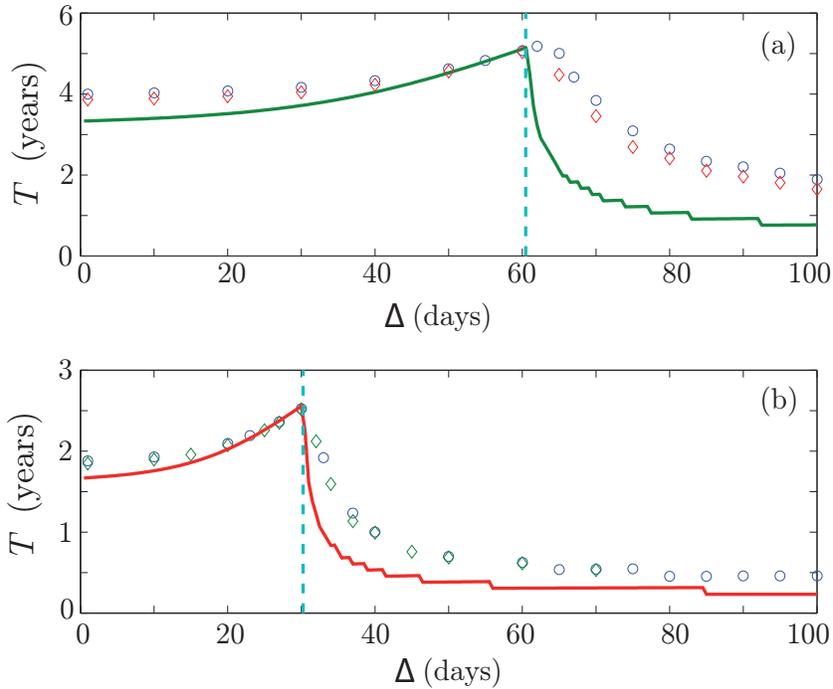,width=11cm}
\caption{Time to transformation obtained from simulations of Eq. (\ref{conc}) with different types of initial data and parameter values $S_f=0.85, U_0 = 0.3, U_* = 0.5$ as a function of the interdose spacing $\Delta$ for (a) $\rho =0.005$ and (b) $\rho = 0.01$. The dashed lines indicate the optimal interdose spacing given by Eq. (\ref{optfin}) for these parameters. Circles corresponds to the results of simulations with symmetric initial data of the form $u_0(x,y,z) = U_0 \text{sech}(0.3 \sqrt{x^2+y^2+z^2})$, while diamonds mark the results of simulations with asymmetric initial data given by $u_0(x,y,z) = 
 0.276 \ \text{sech}(0.3 \sqrt{x^2+y^2+z^2})+ 
        0.184 \ \text{sech}\left(0.3\sqrt{(x-10)^2+(y+8)^2+z^2}\right) + 
        0.184 \ \text{sech}\left(0.3\sqrt{(x+3)^2+(y-1)^2+(z-10)^2}\right),$ that has the same initial value for $U(0) = \max_{x\in \Omega} u(x,y,z) = 0.3$ as in the case of symmetric data. The solid lines correspond to the values obtained from direct simulations of Eqs. (\ref{sim}), that provide upper bounds for the real three-dimensional dynamics. 
 \label{3Dfig}}
\end{center}
\end{figure}

We have computed the dynamics of the amplitude $U_h(t) = \max_{\Omega} u_h(x,t)$, $u_h$ being the discrete approximation to $u$ on the lattice obtained from the three-dimensional simulations. Given a therapeutical schedule, defined by its dose interspacing $\Delta$ we compute the time $T$ that takes the amplitude to reach the critical level for the malignant transformation $U_*$. Fig. \ref{3Dfig} summarizes our results.

First, it is clear from Fig. \ref{3Dfig} that the transformation times obtained from the full numerical simulations of Eqs. (\ref{conc}) are typically larger than those obtained from Eq. (\ref{sim}). This is reasonable since Eq. (\ref{sim}) provides the dynamics of an upper bound for $u(x,t)$ and thus reaches the critical amplitude well before $u(x,t)$. Thus the real gain in days before the malignant transformation occurs is smaller in the framework of Eq. (\ref{conc}) than the one obtained from Eq. (\ref{TD}) (e.g. the ones depicted in Fig. \ref{comparisonT}), but still substantial and about one year for the two sets of parameters studied.

Both the three-dimensional data and the bounds are very close for values of $\Delta$ close to the optimal ones. What it is more interesting is that the value for $\Delta$ obtained from Eq. (\ref{optfin}) predict very accurately the correct optimal values of the optimal dose interspacing found from the full three-dimensional simulations. It is remarkable that such a simple formula can provide the optimal values for 

\section{Discussion and conclusions}
\label{conclusions}

In this paper we have found simple equations providing the optimal dose interspacing in radiotherapy protocols for low-grade gliomas, under the assumptions of equal doses per fraction and fixed spacing between doses. The predictions are obtained within the framework of a simple model but have been found to have an excellent agreement with the numerical results from fully spatio-temporal three-dimensional mathematical models. 

From the point of view of applications, the main conclusion of this paper is that it may be possible to improve the efficacy of radiotherapy as an upfront therapy for the management of low grade gliomas. 

Specifically, the basics for improvement consists on enlarging substantially the interval between doses from the typical choice $\Delta = 1$ day to larger values obtained through Eq. (\ref{optfin}). Thus, the results of Sec. \ref{results} provide a theoretical support for extremely protracted therapies for low grade gliomas. It is interesting to point out that the results have a very clear biological meaning. Since tumor regrowth is known to be faster for small tumor densities, what it is called the \emph{accelerated repopulation} phenomenon, it pays out to leave the tumor density grow while keeping its damage under control. Our logistic growth term captures this phenomenon to some extent.

Indeed, the effect of radiation over a population of tumor cells with so-small fraction of proliferating cells would be minimal and thus a more effective way of using this therapy would be to delay fractions in time to hit new tumor subpopulations once they continue through the cell cycle. These effects, not accounted for in our model, may be incorporated by including two cell subpopulations, one quiescent and another proliferative coupled with the first one, i.e. a compartment of stem cells that in high grade glioma have a crucial role in the response to therapies \citep{Bao2006,Beier2008,Massey2012,Chen2012,Barrett2012,Dirks2001}

A relevant fact is that taking $\Delta$ above the optimal value $\Delta_{\text{opt}}$ leads to a sharp drop of the gain in time to the malignant transformation and even a worse outcome than 
the choice $\Delta =1$ (see Figs. \ref{RTeffect} and \ref{3Dfig}). Thus, to exploit the full power of our approach it may be necessary to have good estimates for  the tumor proliferation $\rho$ and radiosensitivity $S_f$ parameters. In principle it may be possible to estimate those parameters from MRI measurements of the response to a set of doses of radiation \citep{PerezGarcia2014}, thus a two-step strategy may be designed where a first batch of radiation doses (e.g. one-two weeks) are given intensively ($\Delta = 1$) to treat sympthoms and characterize the tumor parameters from the response on MRI and later the remaining set of fractions are given optimally according to the found tumor parameters $\rho, S_f$. Another possibility is to take conservative values of $\Delta \simeq 20-30$ day that would lead to improvements, though non-optimal, for 
 most tumors except for those very aggressive (large $\rho$) and then in case of radiological confirmation of tumor control after some period of time (e.g. one year) enlarge the interval between fractions.
 
In this paper we have tried only to make a first approach to the problem through a proof of concept of the basic idea. Further gains may be possible by solving the full optimization problem were the number of doses $N$, treatment times $t_j$ and individual doses $d_j$ are variables to be choosen to maximize the malignant transformation time (\ref{T2}). While the mathematical problem is interesting and we intend to address it in the future, complex strategies depending too much on the values system parameters might be difficult or even impossible to validate and implement due to parameter uncertainty and may have no impact on the clinical management of these tumors. Thus we have opted for a simpler approach easier to validate in animal models and to translate into current clinical practice. 

Can the extreme protraction concept be exported to other therapies such as chemotherapy for low grade gliomas? Currently, the most used chemotherapeutical scheme consists of a week treatment (an oral dose per day) per month, being repeated for several cycles (months). Although this scheme is already more extended in time, there might be benefit either in enlarging the interval between cycles or in extending the interval between doses within a  given cycle. Since side effects in chemotherapy are systemic instead of mostly local for RT, different models and/or restrictions are to be used for that problem, that will be considered in the future.


We hope our results will motivate research on animal models of low-grade gliomas and a further translation to the clinical practice if sucessful, on the benefit of patients of this lethal disease.

\section*{Acknowledgements}

This work has been partially supported by the Ministerio de Econom\'{\i}a y Competitividad (Spain), under grant MTM2012-31073.
I wish to acknowledge Alicia Mart\'{\i}nez-Gonz\'alez, Juan Belmonte (Universidad de Castilla-La Mancha), Juan M. Sep\'ulveda (Hospital 12 de Octubre), Luis P\'erez-Romasanta (Hospital Universitario de Salamanca), Pilar S\'anchez (Instituto de Salud Carlos III) for discussions.

\end{document}